\documentclass[11pt, a4paper, onecolumn]{article}

\newlength{\imgwidth}
\usepackage{cite}
\usepackage{units}
\usepackage{hyperref}
\hypersetup{
    colorlinks=true,
    linkcolor=blue,
    filecolor=magenta,      
    urlcolor=blue,
    citecolor={blue}
    }

\usepackage{fixltx2e}
\usepackage{pdfpages}
\usepackage[margin=2.8cm]{geometry}
\usepackage{titling}
\usepackage{titlesec}
\titleformat*{\section}{\Large\bfseries\sffamily}
\titleformat*{\subsection}{\large\bfseries\sffamily}
\titleformat*{\subsubsection}{\itshape\subsubsectionfont}

\usepackage[font={small,sf}]{caption}
\newenvironment{sciabstract}{%
\bfseries\sffamily}
{}

\usepackage{graphicx}

\pretitle{\LARGE\bfseries\sffamily\noindent}
\title{The Evolution of Popular Music: USA 1960--2010}
\posttitle{\par\vskip 10pt}

\preauthor{\sffamily\bfseries\noindent}
\author
{Matthias Mauch,$^{1\ast}$ Robert M. MacCallum,$^{2}$ Mark Levy,$^{3}$ Armand M. Leroi$^{2}$
\vskip 10pt
\noindent\small\normalfont\sffamily
$^{1}$School of Electronic Engineering and Computer Science, Queen Mary University of London, E1 4NS. United Kingdom, $^{2}$Division of Life Sciences, Imperial College London, SW7 2AZ. United Kingdom, $^{3}$Last.fm, 5--11 Lavingdon Street, London, SE1 0NZ. United Kingdom} 

\postauthor{}
\predate{}
\postdate{}

\date{}

\begin{document} 




\maketitle

\begin{sciabstract}
\noindent In modern societies, cultural change seems ceaseless. The flux of fashion is especially obvious for popular music. While much has been written about the origin and evolution of pop, most claims about its history are anecdotal rather than scientific in nature. To rectify this we investigate the US Billboard Hot 100 between 1960 and 2010. Using Music Information Retrieval (MIR) and text-mining tools we analyse the musical properties of $\sim$17,000 recordings that appeared in the charts and demonstrate quantitative trends in their harmonic and timbral properties. We then use these properties to produce an audio-based classification of musical styles and study the evolution of musical diversity and disparity, testing, and rejecting, several classical theories of cultural change. Finally, we investigate whether pop musical evolution has been gradual or punctuated. We show that, although pop music has evolved continuously, it did so with particular rapidity during three stylistic ``revolutions'' around 1964, 1983 and 1991. We conclude by discussing how our study points the way to a quantitative science of cultural change.
\end{sciabstract}



\section{Introduction}

The history of popular music has long been debated by philosophers, sociologists, journalists and pop stars \cite{adorno1941popular, adorno1975culture, frith1988music, mauch2011anatomy, negus1996popular, stanley2013yeah}. Their accounts, though rich in vivid musical lore and aesthetic judgements, lack what scientists want: rigorous tests of clear hypotheses based on quantitative data and statistics. 
Economics-minded social scientists studying the history of music have done better, but they are less interested in music than the means by which it is marketed \cite{peterson1975cycles, lopes1992innovation, christianen1995cycles, alexander1996entropy, peterson1996measuring, crain1997economics,  tschmuck2006creativity, klein2010chart}. 
The contrast with evolutionary biology---a historical science rich in quantitative data and models---is striking; the more so since cultural and organismic variety are both considered to be the result of modification-by-descent processes \cite{leroi2006recovery, jan2007memetics, maccallum2012evolution, savage2013toward}. Indeed, linguists and archaeologists, studying the evolution of languages and material culture, commonly apply the same tools that evolutionary biologists do when studying the evolution of species \cite{cavalli1981cultural, boyd1985culture, shennan2009pattern, steele2010evolutionary, mesoudi2011cultural, whiten2012culture}. Inspired by their example, here we investigate the ``fossil record'' of American popular music. We adopt a diachronic, historical approach to ask several general questions: Has the variety of popular music increased or decreased over time? Is evolutionary change in popular music continuous or discontinuous? If discontinuous, when did the discontinuities occur? 

Our study rests on the recent availability of large collections of popular music with associated timestamps, and computational methods with which to measure them \cite{serra2012measuring}. Analysis in traditional musicology and earlier data-driven ethnomusicology \cite{lomax1972evolutionary}, while rich in structure \cite{savage2013toward}, is slow and prone to inconsistencies and subjectivity. Promising diachronic studies on popular music exist, but they either lack scientific rigour \cite{mauch2011anatomy}, focus on technical aspects of audio such as loudness, vocabulary statistics and sequential complexity \cite{serra2012measuring, foster2014sequential}, or are hampered by sample size \cite{schellenberg2012emotional}. 
The present work uniquely combines the power of a big, clearly defined diachronic dataset with the detailed examination of musically meaningful audio features.

\begin{figure}
\begin{center}
		\includegraphics[width=\imgwidth, trim=55 300 110 20, clip=true]{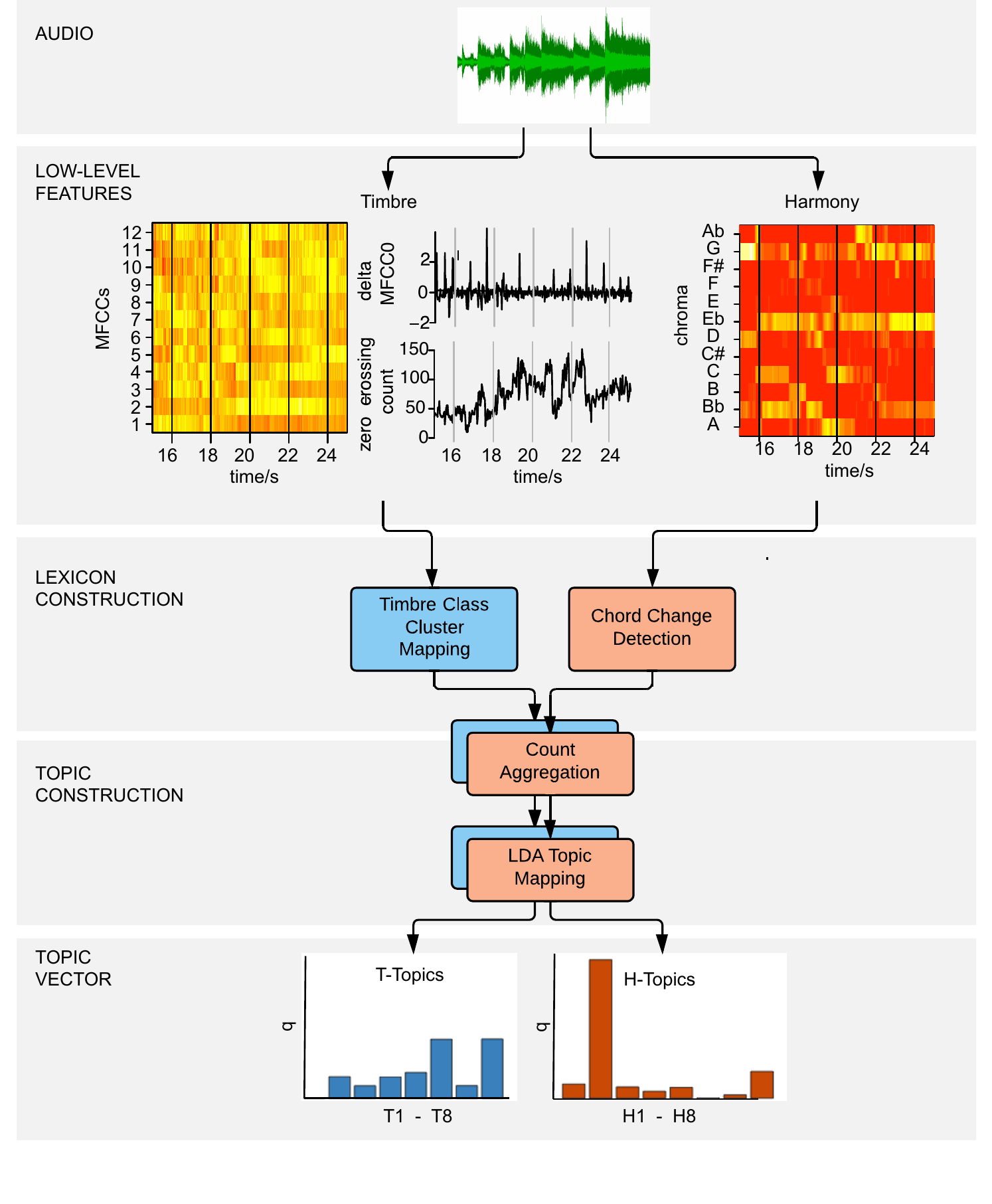}
\end{center}
\caption{\textbf{Data processing pipeline} illustrated with a segment of Queen's \emph{Bohemian Rhapsody}, 1975, one of the few Hot 100 hits to feature an astrophysicist on lead guitar.}
\end{figure}

\begin{figure}
\begin{center}
		\includegraphics[width=\textwidth, trim=10 200 20 30, clip=true]{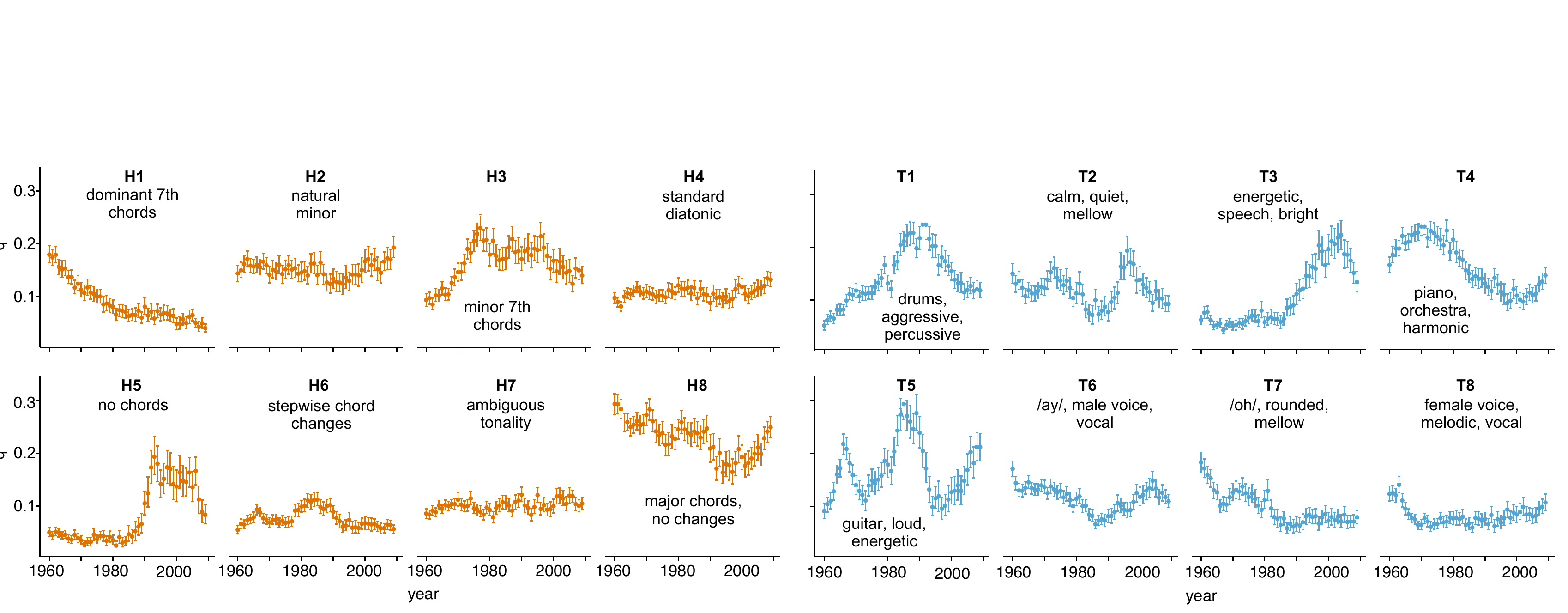}
\end{center}
	\caption{\textbf{Evolution of musical Topics in the Billboard Hot 100.} Mean Topic frequencies ($\bar q$) $\pm$ 95\% CI estimated by bootstrapping.}
\end{figure}

To delimit our sample, we focused on songs that appeared in the US Billboard Hot 100 between 1960 and 2010. We obtained 30-second-long segments of 17,094 songs covering 86\% of the Hot 100, with a small bias towards missing songs in the earlier years. Since our aim is to investigate the evolution of popular taste, we did not attempt to obtain a representative sample of all the songs that were released in the USA in that period of time, but just those that were most commercially successful. To analyse the musical properties of our songs we adopted an approach inspired by recent advances in text mining (figure 1). We began by measuring our songs for a series of quantitative audio features, 12 descriptors of tonal content and 14 of timbre (Supplementary Information M.2--3). These were then discretised into ``words'' resulting in a harmonic lexicon (H-lexicon) of chord changes, and a timbral lexicon (T-lexicon) of timbre clusters (SI M.4). To relate the T-lexicon to semantic labels in plain English we carried out expert annotations (SI M.5). The musical words from both lexica were then combined into 8+8=16 ``Topics'' using Latent Dirichlet Allocation (LDA). LDA is a hierarchical generative model of a text-like corpus, in which every document (here: song) is represented as a distribution over a number of topics, and every topic is represented as a distribution over all possible words (here: chord changes from the H-lexicon, and timbre clusters from the T-lexicon). We obtain the most likely model by means of probabilistic inference (SI M.6). Each song, then, is represented as a distribution over 8 harmonic Topics (H-Topics) that capture classes of chord changes (e.g., ``dominant 7th chord changes'') and 8 timbral Topics (T-Topics) that capture particular timbres (e.g., ``drums, aggressive, percussive'', ``female voice, melodic, vocal'', derived from the expert annotations), with Topic proportions $q$. These Topic frequencies were the basis of our analyses.

\section{Results}

\subsection{The Evolution of Topics}

Between 1960 and 2010, the frequencies of the Topics in the Hot 100 varied greatly: some Topics became rarer, others became more common, yet others cycled (figure 2). To help us interpret these dynamics we made use of associations between the Topics and particular artists as well as genre-tags assigned by the listeners of Last.fm, a web-based music discovery service with $\sim$50m users (electronic supplementary material, M.8). Considering the H-Topics first, the most frequent was H8 (mean $\pm$ 95\%CI: $\bar q=0.236 \pm 0.003$)---major chords without changes. Nearly two-thirds of our songs show a substantial ($>$ 12.5\%) frequency of this Topic, particularly those tagged as \textsc{classic country}, \textsc{classic rock} and \textsc{love} (\href{http://www.eecs.qmul.ac.uk/~matthiasm/descentofpop/fullTables.pdf}{online tables}). Its presence in the Hot 100 was quite constant, being the most common H-Topic in 43 of 50 years. 

Other H-Topics were much more dynamic. Between 1960 and 2009 the mean frequency of H1 declined by about 75\%. H1 captures the use of dominant-7\textsuperscript{th} chords. Inherently dissonant (because of the tritone interval between the third and the minor seventh) these chords are commonly used in Jazz to create tensions that are eventually resolved to consonant chords; in Blues music, the dissonances are typically not resolved and thus add to the  characteristic ``dirty'' colour. Accordingly we find that songs tagged \textsc{blues} or \textsc{jazz} have a high frequency of H1; it is especially common in the songs of Blues artists such as B.B. King and Jazz artists such as Nat ``King'' Cole. The decline of this Topic, then, represents the lingering death of Jazz and Blues in the Hot 100.  

The remaining H-Topics capture the evolution of other musical styles. H3, for example, embraces minor-7th chords used for harmonic colour in funk, disco and soul---this Topic is over-represented in \textsc{funk} and \textsc{disco} and artists like Chic and KC \& The Sunshine Band. Between 1967 and 1977, the mean frequency of H3 more than doubles. H6 combines several chord changes that are a mainstay in modal rock tunes and therefore common in artists with big-stadium ambitions (e.g., M{\"o}tley Cr{\"u}e, Van Halen, REO Speedwagon, Queen, Kiss and Alice Cooper). Its increase between 1978 and 1985, and subsequent decline in the early 1990s, slightly earlier than predicted by the BBC \cite{barfield2007champions}, marks the age of Arena Rock. Of all H-Topics, H5 shows the most striking change in frequency. This Topic, which captures the absence of an identifiable chord structure, barely features in the 1960s and 1970s when, a few spoken-word-music collages aside (e.g., those of Dickie Goodman), nearly all songs had clearly identifiable chords. H5 starts to become more frequent in the late 1980s and then rises rapidly to a peak in 1993. This represents the rise of Hip Hop, Rap and related genres, as exemplified by the music of Busta Rhymes, Nas, and Snoop Dog, who all use chords particularly rarely (\href{http://www.eecs.qmul.ac.uk/~matthiasm/descentofpop/fullTables.pdf}{online tables}). 

The frequencies of the timbral Topics, too, evolve over time. T3, described as ``energetic, speech, bright'', shows the same dynamics as H5 and is also associated with the rise of Hip Hop-related genres. Several of the other timbral Topics, however, appear to rise and fall repeatedly, suggesting recurring fashions in instrumentation. For example, the evolution of T4 (``piano, orchestra, harmonic'') appears sinusoidal, suggesting a return in the 2000s to timbral qualities prominent in the 1970s. T5 (``guitar, loud, energetic'') underwent two full cycles with peaks in 1966 and 1985, heading upward once more in 2009. The second, larger, peak coincides with a peak in H6, the chord-changes also associated with stadium rock groups such as M{\"o}tley Cr{\"u}e (\href{http://www.eecs.qmul.ac.uk/~matthiasm/descentofpop/fullTables.pdf}{online tables}). Finally, T1 (``drums, aggressive, percussive'') rises continuously until 1990 which coincides with the spread of new percussive technology such as drum machines and the gated reverb effect famously used by Phil Collins on \emph{In the air tonight}, 1981. Accordingly, T1 is overrepresented in songs tagged \textsc{dance}, \textsc{disco} and \textsc{new wave} and artists such as The Pet Shop Boys. After 1990, the frequency of T1 declines: the reign of the drum machine was over.

\subsection{The varieties of music}

\begin{figure}[t]
	\begin{center}
	\vspace{-2em}
		\includegraphics[width=\imgwidth, trim=120 620 190 35, clip=true]{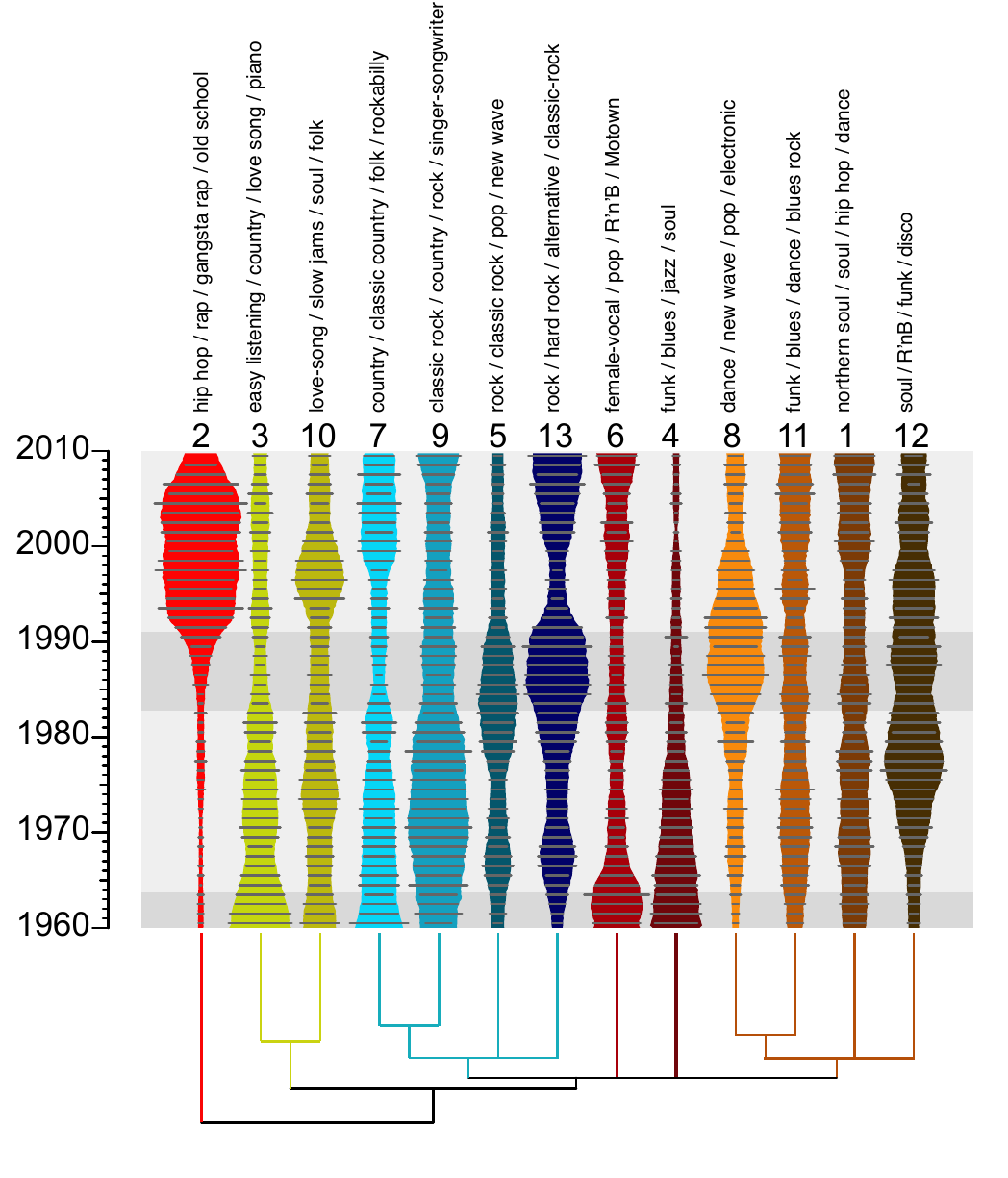}
	\end{center}
	\caption{\textbf{Evolution of musical styles in the Billboard Hot 100.} The evolution of 13 Styles, defined by $k$-means clustering on principal components of Topic frequencies. The width of each spindle is proportional to the frequency of that style, normalised to each year. The spindle contours are based on a $\pm$2-year moving average smoother; unsmoothed yearly frequencies are shown as grey horizontal lines. A hierarchical cluster analysis on the $k$-means centroids grouped our Styles into several larger clusters here represented by a tree: an \textsc{easy-listening + love-song} cluster, a \textsc{country + rock} cluster, and \textsc{soul + funk + dance} cluster; the fourth, most divergent, cluster only contains the \textsc{hip hop + rap}-rich Style 2. All resolved nodes have $\geq 75\%$ bootstrap support.  Labels list the four most highly over-represented Last.fm user tags in each Style according to our enrichment analysis; see electronic supplementary material table~S1 for full results. Shaded regions define eras separated by musical revolutions (figure 5). }
	\label{fig:clustering}
\end{figure}

To analyse the evolution of musical variety we began by classifying our songs. Popular music is classified into genres such as \textsc{country}, \textsc{rock and roll}, \textsc{rhythm and blues} (\textsc{R'n'B}) as well as a multitude of subgenres (\textsc{dance-pop}, \textsc{synthpop}, \textsc{heartland rock}, \textsc{roots rock} etc.). Such genres are, however, but imperfect reflections of musical qualities. Popular music genres such as \textsc{country} and \textsc{rap} partially capture musical styles but, besides being informal, are also based on non-musical factors such as the age or ethnicity of performers  (e.g., \textsc{classic rock} and \ \textsc{K[orean]-Pop}) \cite{negus1996popular}. For this reason we constructed a taxonomy of 13 Styles by $k$-means clustering on principal components derived from our Topic frequencies (figure 3 and electronic supplementary material M.9). We investigated all $k < 25$ and found that the best clustering solution, as determined by mean silhouette score, was $k = 13$.

In order to relate Last.fm tags to the style Style clusters, we used a technique called enrichment analysis from bio-informatics. This technique is usually applied to arrive at biological interpretations of sets of genes, i.e.\ to find out what the ``function'' of a set of genes is. Applying the GeneMerge enrichment-detection algorithm \cite{castillodavis2003genemerge} to our Style data, we found that all Styles are strongly enriched for particular tags, i.e.\ for each Style some Last.fm tags are significantly over-represented (table S1), so we conclude  that they capture at least some of the structure of popular music perceived by consumers. The evolutionary dynamics of our Styles reflect well-known trends in popular music. For example, the frequency of Style 4, strongly enriched for \textsc{jazz},  \textsc{funk},  \textsc{soul} and related tags, declines steadily from 1960 onwards. By contrast, Styles 5 and 13, strongly enriched for \textsc{rock}-related tags, fluctuate in frequency, while Style 2, enriched for \textsc{rap}-related tags, is very rare before the mid-1980s but then rapidly expands to become the single largest Style for the next thirty years, contracting again in the late 2000s. 

What do our Styles represent? Figure 3 shows that Styles and their evolution relate to discrete sub-groups of the charts (genres), and hierarchical cluster analysis suggests that styles can be grouped into a higher hierarchy. However, we suppose that, unlike organisms of different biological species, all the songs in the charts comprise one large, highly structured, meta-population of songs linked by a network of ancestor-descendant relationships arising from songwriters imitating their predecessors \cite{zollo2003songwriters}. Styles and genres, then, represent populations of music that have evolved unique characters (Topics), or combinations of characters, in partial geographic or cultural isolation, e.g., \textsc{country} in the Southern USA during the 1920s or \textsc{rap} in the South Bronx of the 1970s. These Styles rise and fall in frequency over time in response to the changing tastes of songwriters, musicians and producers, who are in turn influenced by the audience.

\subsection{Musical diversity has not declined}

Just as paleontologists have discussed the tempo and mode of evolutionary change in the fossil record \cite{simpson1944tempo}, historians of music have discussed musical change and the processes that drive it. Some have argued that oligopoly in the media industries has caused a relentless decline in cultural diversity of new music \cite{adorno1941popular, adorno1975culture}, while others suggest that such homogenizing trends are periodically interrupted by small competitors offering novel and varied content resulting in ``cycles of symbol production'' \cite{peterson1975cycles, peterson1996measuring}. For want of data there have been few tests of either theory \cite{lopes1992innovation, christianen1995cycles, alexander1996entropy,  tschmuck2006creativity}. 

\begin{figure}
	\begin{center}
		\includegraphics[width=\imgwidth, trim=70 790 225 30, clip=true]{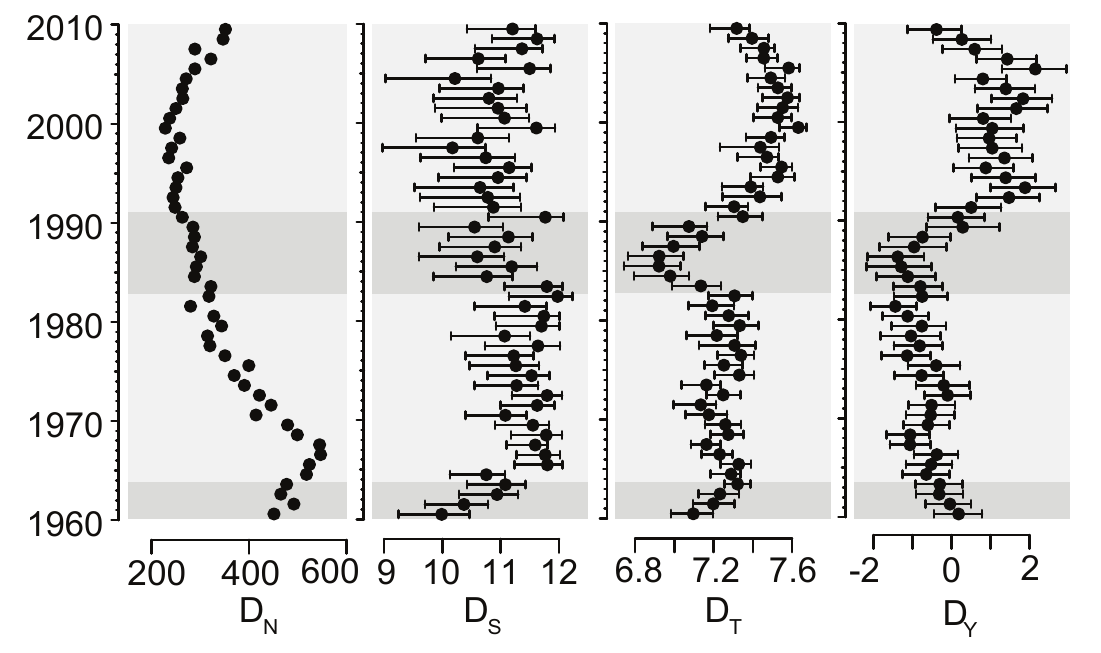}
	\end{center}
	\caption{\textbf{Evolution of musical diversity in the Billboard Hot 100.}  We estimate four measures of diversity. From left to right:  Song number in the charts, $D_{\mathrm{N}}$, depends only on the rate of turnover of unique entities (songs), and takes no account of their phenotypic similarity. Class diversity, D$_{\mathrm{S}}$, is the effective number of Styles and captures functional diversity. Topic diversity, D$_{\mathrm{T}}$, is the effective number of musical Topics used each year, averaged across the Harmonic and Timbral Topics. Disparity, D$_{\mathrm{Y}}$, or phenotypic range is estimated as the total standard deviation within a year. Note that although in ecology D$_{\mathrm{S}}$ and D$_{\mathrm{Y}}$ are often applied to sets of distinct species or lineages they need not be; our use of them implies nothing about the ontological status of our Styles and Topics. For full definitions of the diversity measures see electronic supplementary material, M.11. Shaded regions define eras separated by musical revolutions (figure 5).}
	\label{fig:diversity}
\end{figure}

To test these ideas we estimated four yearly measures of diversity (figure 4).  We found that although all four evolve, two---Topic diversity and disparity---show the most striking changes, both declining to a minimum around 1984, but then rebounding and increasing to a maximum in the early 2000s. Since neither of these measures track song number, their dynamics cannot be due to varying numbers of songs in the Hot 100; nor, since our sampling over 50 years is nearly complete, can they be due to the over-representation of recent songs---the so-called ``pull of the recent'' \cite{jablonski2003impact}. Instead, their dynamics are due to changes in the frequencies of musical styles.

The decline in Topic diversity and disparity in the early 1980s is due to a decline of timbral rather than harmonic diversity (electronic supplementary material, figure S1). This can be seen in the evolution of particular topics (figure 2). In the early 1980s timbral Topics T1 (drums, aggressive, percussive) and T5 (guitar, loud, energetic) become increasingly dominant; the subsequent recovery of diversity is due to the relative decrease in frequency of the these topics as T3 (energetic, speech, bright) increases. Put in terms of Styles, the decline of diversity is due to the dominance of genres such as \textsc{new wave}, \textsc{disco}, \textsc{hardrock}; its recovery is due to their waning with the rise of \textsc{rap} and related genres (figure 2). Contrary to current theories of musical evolution, then, we find no evidence for the progressive homogenisation of music in the charts and little sign of diversity cycles within the 50 year time frame of our study. Instead, the evolution of chart diversity is dominated by historically unique events: the rise and fall of particular ways of making music.

\subsection{Musical evolution is punctuated by revolutions}

\begin{figure}
	\begin{center}
		\includegraphics[width=\imgwidth, trim=40 490 265 50, clip=true]{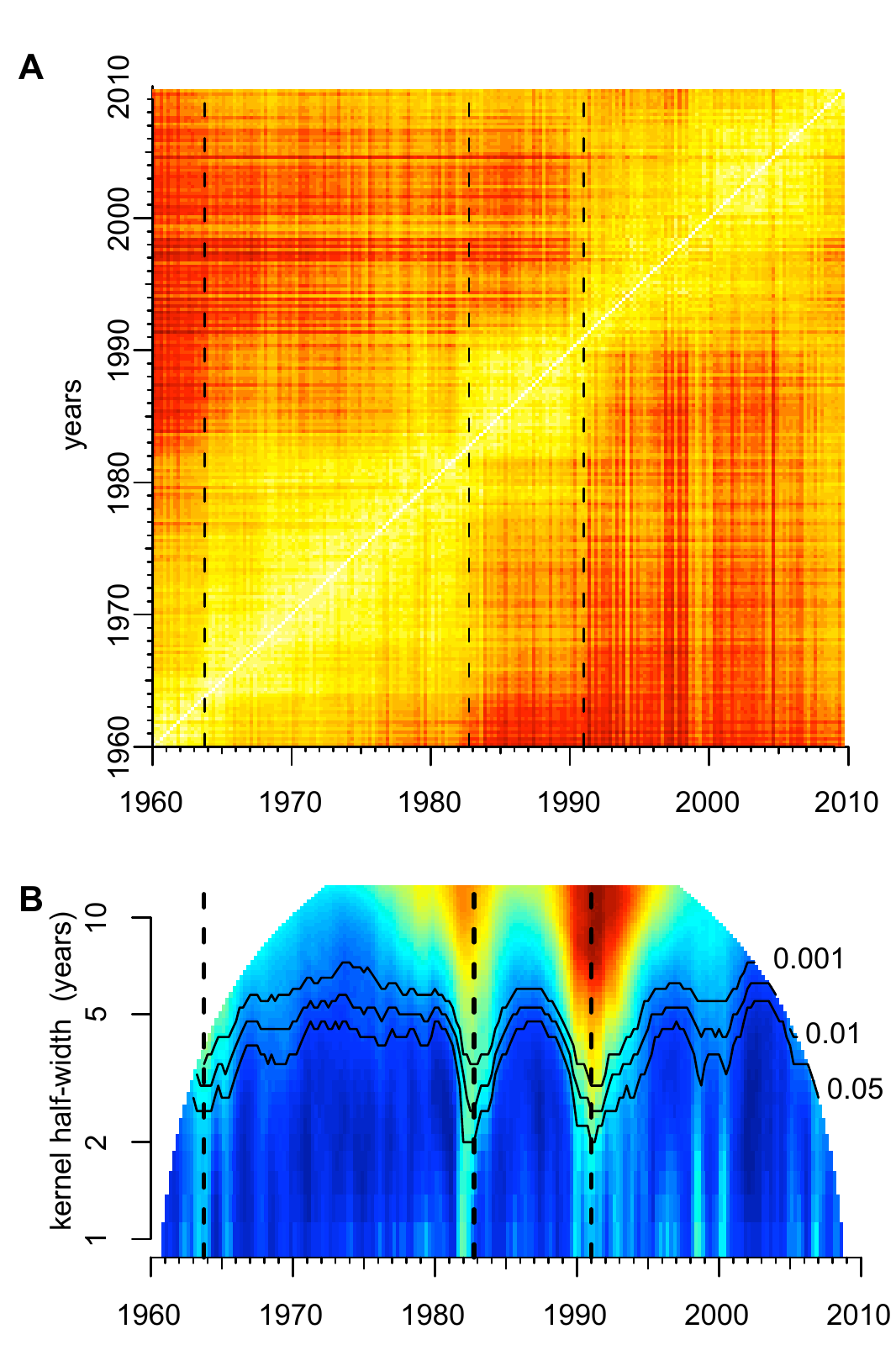}
	\end{center}
	\caption{\textbf{Musical revolutions in the Billboard Hot 100.} \textbf{A}. Quarterly pairwise distance matrix of all the songs in the Hot 100. \textbf{B}. rate of stylistic change based on Foote Novelty over successive quarters for all windows 1--10 years, inclusive. The rate of musical change---slow-to-fast---is represented by the colour gradient blue, green, yellow, red, brown: 1964, 1983, and 1991 are periods of particularly rapid musical change. Using a Foote Novelty kernel with a half-width of 3 years results in significant change in these periods, with Novelty peaks in 1963--Q4 ($P<0.01$), 1982--Q4 ($P<0.01$) and 1991--Q1 ($P<0.001$) marked by dashed lines. Significance cutoffs for all windows were empirically determined by random permutation of the distance matrix. Significance contour lines with $P$ values are shown in black.}
	\label{fig:revolutions}
\end{figure}

The history of popular music is often seen as a succession of distinct eras, e.g., the ``Rock Era'', separated by revolutions \cite{frith1988music, stanley2013yeah, tschmuck2006creativity}. Against this, some scholars have argued that musical eras and revolutions are illusory \cite{negus1996popular}. Even among those who see discontinuities, there is little agreement about when they occurred. The problem, again, is that data have been scarce and objective criteria for deciding what constitutes a break in a historical sequence, scarcer yet. 

To determine directly whether rate discontinuities exist we divided the period 1960--2010 into 200 quarters and used the principal components of the Topic frequencies to estimate a pairwise distance matrix between them (figure 5A). This matrix suggested that, while musical evolution was ceaseless, there were periods of relative stasis punctuated by periods of rapid change. To test this impression we applied a method from Music Information Retrieval, Foote Novelty, which estimates the magnitude of change in a distance matrix over a given temporal window \cite{foote2000automatic}. The method relies on a kernel matrix with a checkerboard pattern. Since a distance matrix exposes just such a checkerboard pattern at change points \cite{foote2000automatic}, convolving it with the checkerboard kernel along its diagonal directly yields the novelty function (SI M.11). We calculated Foote Novelty for all windows between 1 and 10 years and, for each window, determined empirical significance cutoffs based on random permutation of the distance matrix. We identified three revolutions: a major one around 1991 and two smaller ones around 1964 and 1983 (figure 5B). From peak to succeeding trough, the rate of musical change during these revolutions varied 4- to 6-fold. 

This temporal analysis, when combined with our Style clusters (figure 3), shows how musical revolutions are associated with the expansion and contraction of particular musical styles. Using quadratic regression models, we identified the Styles that showed significant ($P<0.01$) change in frequency against time in the six years surrounding each revolution (electronic supplementary material, table S2). We also carried out a Style-enrichment analysis for the same periods (electronic supplementary material, table S2). Of the three revolutions 1964 was the most complex, involving the expansion of several Styles---1, 5, 8, 12 and 13---enriched at the time for \textsc{soul} and \textsc{rock}-related tags. These gains were bought at the expense of Styles 3 and 6 both enriched for \textsc{doowop} among other tags. The 1983 revolution is associated with an expansion of three Styles---8,11 and 13---here enriched for \textsc{new wave}, \textsc{disco} and \textsc{hard rock}-related tags and the contraction of three Styles---3, 7 and 12---here enriched for \textsc{soft rock}, \textsc{country}-related or \textsc{soul + r'n'b}-related tags. The largest revolution of the three, 1991, is associated with the expansion of Style 2, enriched for \textsc{rap}-related tags, at the expense of Styles 5 and 13, here enriched for \textsc{rock}-related tags.  The rise of  \textsc{rap} and related genres appears, then, to be the single most important event that has shaped the musical structure of the American charts in the period that we studied. 

\subsection{The British did not start the American revolution of '64}
\begin{figure}
	\begin{center}
		\includegraphics[width=\imgwidth, trim=10 100 150 30, clip=true]{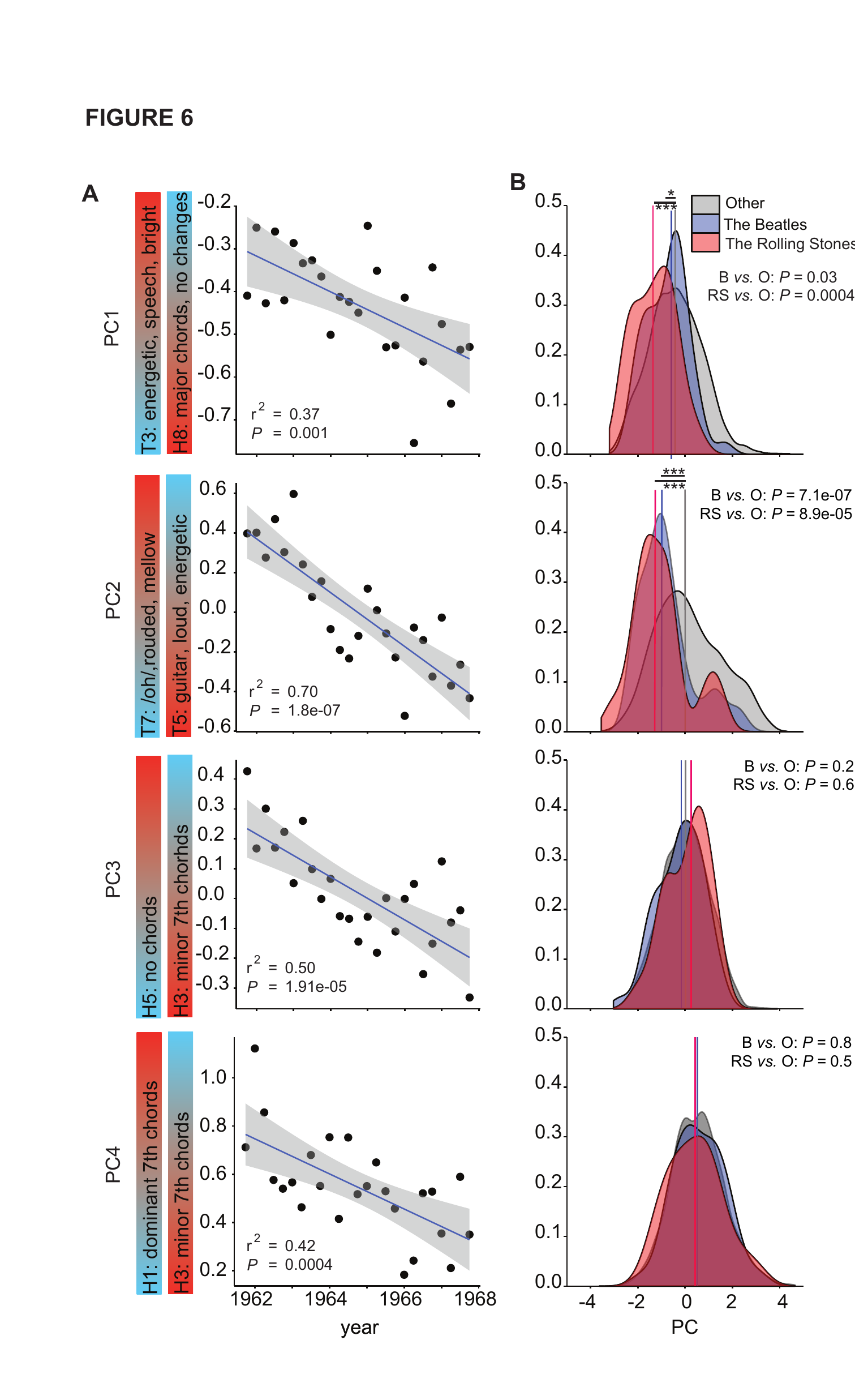}
	\end{center}
	\caption{\textbf{The British Invasion in the American revolution of 1964.} Top to Bottom: PC1--PC4.  \textbf{A}. Linear evolution of quarterly medians of four PCs in the six years (24 quarters) flanking 1963--Q4, the peak of the 1964 revolution. The population medians of all four PCs decrease, and these decreases begin well before the start of the British Invasion in late 1963, implying that BI acts cannot be solely responsible for the changes in musical style evident at the time. For each PC, the two topics that load most strongly are indicated, with sign of correlation---high, red to low, blue---indicated (electronic supplementary material, figure S2).  \textbf{B}. Frequency density distributions of four PCs for Beatles, Rolling Stones and songs by all other artists around the 1964 revolution. For PC1 and PC2, but not PC3 and PC4, The Beatles and The Rolling Stones have significantly lower median values than the rest of the population, indicated with arrows, implying that these BI artists adopted a musical style that exaggerated existing trends in the Hot 100 towards increased use of major chords and decreased use of ``bright'' speech (PC1) and increased guitar-driven aggression and decreased use of mellow vocals (PC2). Vertical lines represent medians; $P$ values based on Mann-Whitney-Wilcoxon rank sum test; The Beatles (B): $N$ = 46; The Rolling Stones (RS): $N$ = 20; Other artists (O): $N$ = 3114.} 
	\label{fig:BI}
\end{figure}

Our analysis does not reveal the origins of musical styles; rather, it shows when changes in style frequency affect the musical structure of the charts. Bearing this in mind we investigated the roles of particular artists in one revolution. On 26 December, 1963, The Beatles released \emph{I want to hold your hand} in the USA. They were swiftly followed by dozens of British acts who, over the next few years, flooded the American charts.  It is often claimed that this ``British Invasion" (BI) was responsible for musical changes of the time \cite{fitzgerald1995when}. Was it? As noted above, around 1964 many Styles were changing in frequency; many principal components of the Topic frequencies show linear changes in this period too. Inspection of the first four PCs shows that their evolutionary trajectories were all established before 1964, implying that, while the British may have contributed to this revolution, they could not have been entirely responsible for it (figure 6A). We then compared two of the most successful BI acts, The Beatles and The Rolling Stones, to the rest of the Hot 100 (figure 6B). In the case of PC1 and PC2, the songs of both bands have (low) values that anticipate the Hot 100's trajectory: for these musical attributes they were literally ahead of the curve. In the case of PC3 and PC4 their songs resemble the rest of the Hot 100: for these musical attributes they were merely on-trend.  Together, these results suggest that, even if the British did not initiate the American revolution of 1964, they did exploit it and, to the degree that they were imitated by other artists, fanned its flames. Indeed, the extraordinary success of these two groups---66 Hot 100 hits between them prior to 1968---may be attributable to their having done so.

\section{Discussion and Conclusions}

Our findings provide a quantitative picture of the evolution of popular music in the USA over the course of fifty years. As such, they form the basis for the scientific study of musical change. Those who wish to make claims about how and when popular music changed can no longer appeal to anecdote, connoisseurship and theory unadorned by data. Similarly, recent work has shown that it is possible to identify discrete stylistic changes in the history of Western classical music by clustering on motifs extracted from a corpus of written scores \cite{rodriguez2013perceptual}. Insofar that our approach is based on audio, it can also be applied to music for which no scores exist, including that from pre-Modern cultures \cite{savage2013toward, lomax1968folk, lomax1972evolutionary}. We have already applied a similar approach to the classification of Art music (``classical music'') into historical periods \cite{weiss2014timbre}. More generally, music is a natural starting point for the study of stylistic evolution because it is not only a universal human cultural trait \cite{brown1991human}, but also measurable, largely determined by form, and available in a relatively standardised format (digital recordings).

Our study is limited in several ways. First, it is limited by the features studied. Our measures must capture only a fraction of the phenotypic complexity of even the simplest song; other measures may give different results. However, the finding that our classifications are supported by listener genre-tags gives us some confidence that we have captured an important part of the perceptible variance of our sample. Second, in confining our study to the Hot 100, 1960--2010, we have only sampled a small fraction of the new singles released in the USA; a complete picture would require compiling a database of several million songs, which in itself is a challenge \cite{bertin2011million}. Given that the Hot 100 is certainly a biased subset of these songs, our conclusions cannot be extended to the population of all releases. Finally, we are interested in extending the temporal range of our sample to at least the 1940s---if only to see whether 1955 was, as many have claimed, the birth date of Rock'n'Roll \cite{peterson1990why}.

We have not addressed the causes of the dynamics that we detect. Like any cultural artefact---and any living organism---music is the result of a variational-selection process \cite{leroi2006recovery, jan2007memetics, maccallum2012evolution, savage2013toward}. In evolutionary biology, causal explanations of organismal diversity appeal to intrinsic constraints (developmental or genetic), ecological factors (competition among individuals or lineages) and stochastic events (e.g., rocks from space) \cite{erwin2007disparity, gould2002structure, jablonski2008species}. By analogy, a causal account of the evolution of music must ultimately contain an account of how musicians imitate, and modify, existing music when creating new songs, that is, an account of the mode of inheritance, the production of musical novelty, and its constraints. The first of these---inheritance and its constraints---is obscure \cite{pachet2006creativity,mcintyre2008creativity}; the second---selection---less so. The selective forces acting upon new songs are at least partly captured by their rise and fall through the ranks of the charts. Many anecdotal histories of music attempt to explain these dynamics. For example, the rise of rap in the charts has been credited to the television show \emph{Yo, MTV Raps!} first broadcast in 1988 \cite{george2005hiphop}. A general, multilevel, selection theory, not restricted to Mendelian inheritance, should provide a means for such hypotheses to be tested \cite{price1970selection, price1972extension, frank1995george}. 

Finally, we note that the statistical tools used in this study are quite general. Latent Dirichlet Allocation can be used to study the evolving structure of many kinds of assemblages; Foote Novelty can be used to detect rate discontinuities in temporal sequences of distances based on many kinds of phenotypes. Such tools, and the existence of large digital corpora of cultural artefacts---texts, music, images, computer-aided design (CAD) files---now permits the evolutionary analysis of many dimensions of modern culture. We anticipate that the study of cultural trends based upon such datasets will soon constrain and inspire theories about the evolution of culture just as the fossil record has for the evolution of life \cite{michel2011quantitative}.

\section*{Data accessibility}

All methods and supplementary figures and tables are available in the electronic supplementary materials. Extensive data, including song titles, artists, topic frequencies and tags are available from the Figshare repository \href{http://figshare.com/s/df5dd714b62111e4bcc906ec4bbcf141}{main data frame}, \href{http://figshare.com/s/7916c680b62311e48bbf06ec4b8d1f61}{secondary (tag) data frame}. {\bf [TEMPORARY LINKS, WILL BE UPDATED UPON ACCEPTANCE]}
 
 \section*{Authors' contribution}
 
ML provided data; MM, RMM and AML analysed the data; MM and AML conceived of the study, designed the study, coordinated the study and wrote the manuscript. All authors gave final approval for publication.

 \section*{Funding} 
 
 Matthias Mauch is funded by a Royal Academy of Engineering Research Fellowship. 
 
 \section*{Acknowledgments} 
 
 We thank the public participants in this study; Austin Burt, Katy Noland and Peter Foster for comments on the manuscript; Last.fm for musical samples; Queen Mary University of London, for the use of high-performance computing facilities.

\bibliography{../../referencesmm,../../mypapers}
\bibliographystyle{vancouver}

\newpage



\section*{Supplementary material}

# The evolution of popular music: USA 1960-2010: Supporting Information

Matthias Mauch, Robert M. MacCallum, Mark Levy, Armand M. Leroi

15 October, 2014

# Contents

1

# M  Materials and Methods

## M.1  The origin of the songs

Metadata on the complete Billboard Hot 100 charts were obtained through the (now defunct) Billboard API, consisting of artist name, track name, and chart position in every week of the charts from 1957 to early 2010. We use only songs from 1960 through 2009 since these years have complete coverage. Using a proprietary matching procedure, we associated Last.fm MP3 audio recordings with the chart entries. Each recording is 30 seconds long. We use 17,094 songs, covering 86% of the weekly Billboard charts (84% before 2000, 95% from 2000 onward (figure M1). This amounts to 69% of unique audio recordings. The total duration of the music data is 143 hours.

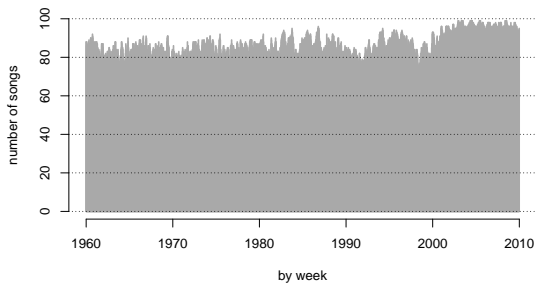

Figure M1: Coverage of the Billboard Hot 100 Charts by week.

To validate our impression that data quality was good, a random sub-sample of 9928 songs was vetted by hundreds of volunteers recruited on the internet. The participants were presented with two recordings, and for each were asked to to answer the question "Does recording [...] have very poor audio quality?". We analysed those 5593 recordings that were judged at least twice. A recording was considered poor quality if it was marked as such by a majority vote. Overall, this was the case in only 3.8% of the recordings, with a bias towards worse quality recordings in the 1960s (9.1%; 1970 and later: 1.8%). To examine the effect of bad songs, we removed them and compared the estimated mean $q$ of each topic (Section M.6) for the total population versus the population of 'good' songs for each year of the 1960s. In no case did we find that they were significantly different. We conclude that recording quality will have a negligible effect on our results.

All songs were decoded to PCM WAV format (44100 Hz, 16 bit). The songs were then band-pass-filtered using the Audio Degradation Toolbox [1] to reduce differences in recording equalisation in the bass and high treble frequencies (stop-band frequencies: 67 Hz, 6000 Hz).

## M.2  Measuring Harmony.

The harmony features consist of 12-dimensional chroma features (also: pitch class profiles) [2]. Chroma is widely used in MIR as a robust feature for chord and key detection [3], audio thumbnailing [4], and automatic structural segmentation [5]. In every frame chroma represents the activations (i.e. the strength)

$$\mathbf{c} = (c_1, \ldots, c_{12})$$

corresponding to the 12 pitch classes in the chromatic musical scale (i.e. that of the piano): A, B♭, B C, ..., G, A♭. We use the NNLS Chroma implementation [6] to extract chroma at the same frame rate as the timbre features (step size: 1024 samples = 23ms, i.e. 43 per second), but with the default frame size

of 16384 samples. The chroma representation (often called chromagram) of the complete 30 s excerpt of "Bohemian Rhapsody" is shown in figure 1 (main text).

## M.3 Measuring Timbre.

The timbre features consist of 12 Mel-frequency cepstral coefficients (MFCCs), one delta-MFCC value, and one Zero-crossing Count (ZCC) feature. MFCCs are spectral-domain audio features for the description of timbre and are routinely used in speech recognition [7] and Music Information Retrieval (MIR) tasks [8]. For every frame, they provide a low-dimensional parametrisation of the overall shape of the signal's Mel-spectrum, i.e. a spectral representation that takes into account human near-logarithmic perception of sound in magnitude (log-magnitude) and frequency (Mel scale). We use the first 12 MFCCs (excluding the $0^{\text{th}}$ component) and additionally one delta-MFCC, calculated as the difference between any two consecutive values of the $0^{\text{th}}$ MFCC component. The MFCCs were extracted using a plugin from the Vamp library (seen 27.03.2014) with the default parameters (block size: 2048 samples = 46ms, step size: 1024 samples = 23ms). This amounts to $\approx$ 43 frames per second. The ZCC (also: zero-crossing rate, ZCR) is a time-domain audio feature which has been used in speech recognition [9] and has been applied successfully to discern drum sounds [10]. It is calculated by simply counting the number of times consecutive samples in a frame are of opposite sign. ZCC is high for noisy signals and transient sounds at the onset of consonants and percussive events. To extract the ZCC we also used a Vamp plugin, extracting features at the same frame rate (43 per second, step size: 1024 sampes = 23ms), but with a block size of 1024 samples. MFCCs and zero crossing counts of "Bohemian Rhapsody" are shown in figure 1 (main text).

## M.4 Making musical lexica

Since we aim to apply topic models to our data (see Section M.6), we need to discretise our raw features into musical lexica. We have one timbral lexicon (T-Lexicon) and one harmonic lexicon (H-Lexicon).

**Timbre.**

In order to define the T-Lexicon we followed an unsupervised feature learning approach by quantising the feature space into 35 discrete classes as follows. First, we randomly selected 20 frames from each of 11350 randomly selected songs (227 from every year), a total of 227,000 frames. The features were then standardised, and de-correlated using principal component analysis (PCA). The PCA components were once more standardised. We then applied model-based clustering (Gaussian mixture models, GMM) to the standardised de-correlated data, using the built-in Matlab function gmdistribution.fit with full covariance matrix [11]. The GMM with 35 mixtures (clusers) was chosen as it minimised the Bayes Information Criterion. We then transformed all songs according to the same PCA, scaling and cluster mapping transformations. In particular, every audio frame was assigned to its most likely cluster according to the GMM. Frames with cluster probabilities of $< 0.5$ were removed.

**Harmony.**

Our H-Lexicon consists of all 192 possible changes between the most frequently used chord types in popular music [12]: major (M), minor (m), dominant 7 (7) and minor 7 chords (m7). We use chord changes because they offer a key-independent way of describing the temporal dynamics of harmony. As a chord is defined by its root pitch class (A,B♭,B,C,...,A♭) and its type, our system gives rise to $4 \times 12 = 48$ chords. Each of the chords can be represented as a binary chord template with 12 elements corresponding to the twelve pitch classes. For example, the four chords with root A are these.

$$CT_{AM} = (1, 0, 0, 0, 1, 0, 0, 1, 0, 0, 0, 0)$$

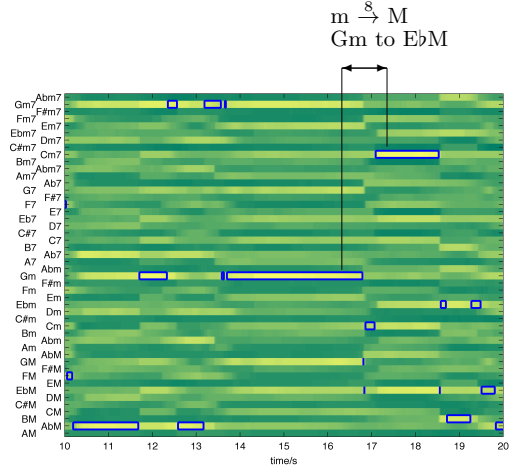

Figure M2: Chord activation, with the most salient chords at any time highlighted in blue. Excerpt of "Bohemian Rhapsody" by Queen.

$$CT_{Am} = (1,0,0,1,0,0,0,1,0,0,0,0)$$
$$CT_{A7} = (1,0,0,0,1,0,0,1,0,0,1,0)$$
$$CT_{Am7} = (1,0,0,1,0,0,0,1,0,0,1,0)$$

At every frame we estimate the locally most likely chord by correlating the chroma vectors (Section ??) with the binary chord templates (see, e.g. [13]), i.e. a given chroma frame $\mathbf{c} = (c_1, \ldots, c_{12})$ is correlated to a chord template $CT = (CT_1, \ldots, CT_{12})$ by using Pearson's rho,

$$\rho_{CT,\mathbf{c}} = \sum_{i=1}^{12} \frac{(CT_i - \overline{CT})(c_i - \overline{\mathbf{c}})}{\sigma_{CT}\sigma_{\mathbf{c}}},$$

where $\bar{\cdot}$ denotes the sample mean and $\sigma_{\cdot}$ denotes the sample standard deviation of the corresponding vector. To smooth these correlation scores over time, we apply a median filter of length 43 (1 second). An example of the resulting smoothed chord activation matrix is shown in figure M2. We then choose the chord with the highest median-smoothed correlation and combine the two chord labels spaced 1 second apart into one chord change label, retaining only the relative root positions of the chords and both chord types [14], as demonstrated in figure M2. If the chord change was ambiguous (mean correlation of the two chords $< 0.4$), the chord change label was set to an additional 193$^{\text{rd}}$ label *NA*.

In summary, we have obtained two lexica of frame-wise discrete labels, one for timbre (35 classes) and one for harmony (193 chord changes). Each allows us to describe a piece of music as a count vector giving counts of timbre classes and chord changes, respectively.

## M.5 Semantic lexicon annotation

Since we can now express our music in terms of lexica of discrete items, we can attach human-readable labels to these items. In the case of the 193 chord changes (H-Lexicon), an intrinsic musical interpretation exists. The most frequent chord changes are given in Table ??, along with some explanations and counts over the whole corpus.

The 35 classes in the T-Lexicon do not have *a priori* interpretations, so we obtained human annotations on a subset of our data. First, we randomly selected 100 songs, two from each year, and concatenated

the audio that belonged to the same of the 35 sound classes from our T-Lexicon using an overlap-add approach. That is, each audio file contained frames from only one of the timbre classes introduced in Section M.4, but from up to 100 songs. The resulting 35 sound class files can be accessed on SoundCloud[1]). We noticed that each of the files does indeed have a timbre characteristic; some captured a particular vowel sound, others noisy hi-hat and crash cymbal sounds, others again very short, percussive sounds. We then asked ten human annotators to individually describe these sounds. Each annotator listened to all 35 files and, for each, subjectively chose 5 terms that described the sound from a controlled vocabulary consisting of the following 34 labels manually compiled from initial free-vocabulary annotations:

*mellow, aggressive, dark, bright, calm, energetic, smooth, percussive, quiet, loud, harmonic, noisy, melodic, rounded, harsh, vocal, instrumental, speech, instrument: drums, instrument: guitar, instrument: piano, instrument: orchestra, instrument: male voice, instrument: female voice, instrument: synthesiser, 'ah', 'ay', 'ee', 'er', 'oh', 'ooh', 'sh', 'ss', [random – I find it hard to judge].*

On average, the most agreed-upon label per class was chosen by 7.5 (mean) of the 10 annotators, indicating good agreement. Even the second- and third-ranking labels were chosen by more than half of the annotators (means 6.4 and 5.68). Figure M3 shows the agreement of the top labels from rank 1 to 8.

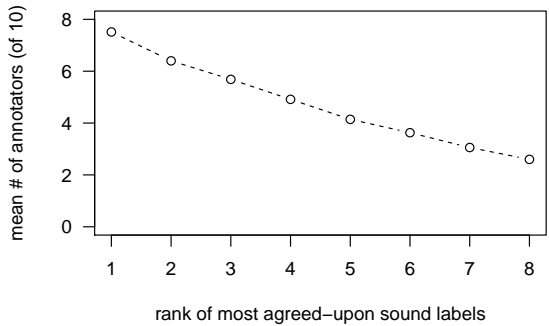

Figure M3: Agreement of the 10 annotators in the semantic sound annotation task.

## M.6 Topic extraction

For timbre and harmony separately, a topic model is estimated from the song-wise counts, using the implementation of Latent Dirichlet Allocation (LDA) [15] provided in the `topicmodels` library [16] for R. LDA is a hierarchical generative model of a corpus. The original model was formulated in the context of a text corpus in which

a) every document (here: song) is represented as a discrete distribution over $N_T$ topics

b) every every topic is represented as a discrete distribution over all possible words (here: H-Lexicon or T-Lexicon entries)

Since the T- and H-Lexicon count vectors introduced in Section M.4 are of the same format as word counts, we can apply the same modelling procedure. That is, by means of probabilistic inference on the model, the LDA method estimates the topic distributions of each song (probabilities of a song using a particular topic) and the topics' lexical distribution (probabilities over the H- and T-lexica) from the lexicon count vectors.

We used the LDA function, which implements the variational expectation-maximization (VEM) algorithm to estimate the parameters, setting the number of topics to 8. Hence, we obtained one model with

---
[1] https://soundcloud.com/descent-of-pop/sets/cluster-sounds

5

8 T-Topics, and one with 8 H-Topics. Topic models allow us to encode every song as a distribution over T- and H-Topics,

$$\mathbf{q}^T = (q_1^T, q_2^T, \ldots, q_8^T)$$
$$\mathbf{q}^H = (q_1^H, q_2^H, \ldots, q_8^H)$$

The probabilities can be interpreted as the proportion of frames in the song belonging to the topic. When it is clear from the context which T- or H-Topic we are concerned with we denote these by the letter $q$, and their mean over a group of songs by $\bar{q}$. Mean values by year for all topics are shown in figure **??** in the main text with 95% confidence intervals based on quantile bootstrapping.

In the same manner, we calculate means and bootstrap confidence intervals for all artists with at least 10 chart entries and all Last.fm tags (introduced in Section M.8) with at least 200 occurrences. The artists with the highest and lowest mean $q$ of each topic and the respective listing of tags can be found online.

## M.7  Semantic topic annotations

In this section we show how to map the semantic interpretations of our harmony and timbre lexica (see Section M.5) to the 8 T-Topics and 8 H-Topics. This allows us to work with the topics rather than the large number of chord changes and sound classes.

**Harmony.**

Each H-Topic is defined as a distribution $P(E_i^H)$ over all H-lexicon entries $E_i^H$, $i = 1, \ldots, 193$ (the 193 different chord changes). The lower half of Table **??** shows the 10 most probable chord changes for each topic with those that have $P(E_i) > 0.01$ emphasised in bold. For example, the most likely chord change in H-Topic 4 is a Major chord followed by another Major chord 7 semitones higher, e.g. C to G. The interpretation of a topic, then, is the coincidence of such chord changes in a piece of music. Interpretations of the 8 H-Topics can be found in Table 1.

**Timbre.**

In order to obtain interpretations for the T-Topics we map the semantic annotations of the T-lexicon (Section M.5) to the topics. The semantic annotations of the T-lexicon come as a matrix of counts $W^* = (w_{ij}^*)$ of annotation labels $j = 1, \ldots, 34$ for each of the sound classes $i = 1, \ldots, 35$. We first normalise the columns $w_{\cdot,j}^*$ by root-mean-square normalisation to obtain a scaled matrix $W_{ij}$ with the elements

$$w_{ij} = \frac{w_{ij}^*}{\sqrt{(1/34) \sum_i (w_{ij}^*)^2}}. \tag{1}$$

The matrix $W = (w_{ij})$ expresses the relevance of the $j^\text{th}$ label for the $i^\text{th}$ sound class. Since T-Topics are compositions of sound classes, we can now simply map these relevance values to the topics by multiplication. The weight $L_j$ of the $j^\text{th}$ label for a T-Topic in which sound class $E_i^T$ appears with probability $P(E_i^T)$ is

$$L_j = \sum_{i=1}^{35} w_{ij} P(E_i^T). \tag{2}$$

The top 3 labels for each T-Topic can be found in Table 1.

Table 1: Topic interpretations.

| | harmonic topics |
|---|---|
| H1 | changes involving dominant 7th chords |
| H2 | natural minor key changes |
| H3 | changes involving minor 7th chords |
| H4 | simple diatonic changes used in major keys |
| H5 | unrecognised changes or no chordal content |
| H6 | stepwise changes indicating modal harmony |
| H7 | ambiguous major/minor attribution |
| H8 | sustained major chords |

| | timbral topics |
|---|---|
| T1 | drums, aggressive, percussive |
| T2 | calm, quiet, mellow |
| T3 | energetic, speech, bright |
| T4 | piano, orchestra, harmonic |
| T5 | guitar, loud, energetic |
| T6 | ay, male voice, vocal |
| T7 | oh, rounded, mellow |
| T8 | female voice, melodic, vocal |

## M.8 User-generated tags

The Last.fm recordings are also associated with tags, generated by Last.fm users, which we obtained via a proprietary process. The tags are usually genre-related (POP, SOUL), but a few also contain information about the instrumentation, feel (PIANO, SUMMER), references to particular artists and others. We removed references to particular artists and joined some tags that were semantically identical. After the procedure we had tags for 16085 (94%) of the songs, with a mean tag count of 2.7 per song (median: 3, mode: 4).

## M.9 Identifying musical Styles clusters: $k$-means and silhouette scores

In order to identify musical styles from our data measurements, we first used the $17094 \times 16$ (i.e. songs $\times$ topics) data matrix of all topic probabilities $\mathbf{q}^T$ and $\mathbf{q}^H$, and de-correlated the data using PCA (see also figure S2). The resulting data matrix has 14 non-degenerate principal components, which we used to cluster the data using $k$-means clustering. We chose a cluster number of 13 based on analysing of the mean silhouette width [17] over a range of $k = 2, \ldots, 25$ clusters, each started with 50 random initialisations. The result of the best clustering at $k = 13$ is chosen, and each song is thus classified to a style $s \in \{1, \ldots, 13\}$ (figure M4).

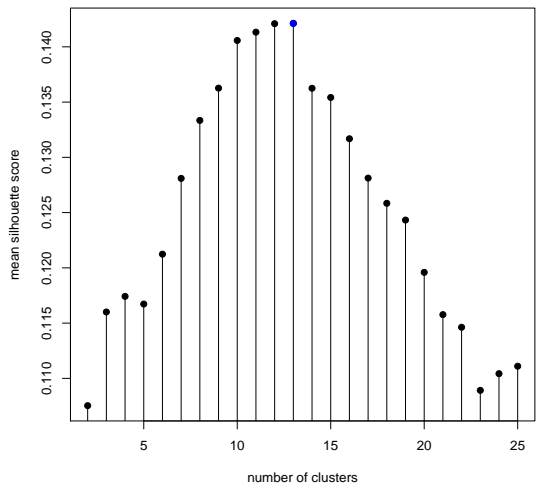

Figure M4: Mean silhouette scores. The optimal number of clusters, $k = 13$ is highlighted in blue.

## M.10 Diversity metrics

In order assess the diversity of a set of songs (usually the songs having entered the charts in a certain year) we calculate four different metrics: number of songs ($D_N$), effective number of styles (style diversity, $D_C$), effective number of topics (topic diversity, $D_T$) and disparity (total standard deviation, $D_Y$). The following paragraphs explain these metrics.

**Number of songs.**

The simplest measure of complexity is the number of songs $D_N$. We use it to show that other diversity metrics are not intrinsically linked to this measure.

**Effective number of Styles.**

In the ecology literature, diversity refers to the effective number of species in an ecosystem. Maximum diversity is achieved when the species' frequencies are all equal, i.e. when they are uniformly distributed. Likewise, minimum diversity is assumed when all organisms belong to the same species. According to [18], diversity for a population of $N_s$ species can formally be defined as

$$D_S = \exp\left(-\sum_{i=1}^{N_s} s_i \ln s_i\right) \quad (3)$$

where $s_i$, $i = 1, \ldots, N_s$ represents the relative frequency distribution over the $N_s$ species such that $\sum_i s_i = 1$. In particular, the maximum value assumed when all species' relative frequencies are equal is $D = N_s$. If, on the other hand, only one species remains, and all others have frequencies of zero, then $D = 1$, the minimum value.

We use this exact definition to describe the year-wise diversity of acoustical Style clusters in our data (recall that each song has only one Style, but a mixture of Topics). For every year we calculate

the proportion songs $s_i$, $i = 1, \ldots, 13$ belonging to each of the 13 Styles, and hence we use $N_s = 13$ to calculate $D_S \in [1, 13]$.

**Effective number of Topics.**

The probability $q$ of a certain topic in a song (see Section M.6) provides an estimate of the proportion of frames in a song that belong to that topic. By averaging over the year, we can get an estimate of the proportion $\bar{q}$ of frames in the whole year, i.e. for all T- and H-Topics we obtain the yearly measurements

$$\bar{\mathbf{q}}^T = (\bar{q}_1^T, \bar{q}_2^T, \ldots, \bar{q}_8^T)$$
$$\bar{\mathbf{q}}^H = (\bar{q}_1^H, \bar{q}_2^H, \ldots, \bar{q}_8^H).$$

Figuratively, we throw all audio frames of all songs into one big bucket pertaining to a year, and estimate the proportion of each topic in the bucket. From these yearly estimates of topic frequencies we can now calculate the effective number of T- and H-Topics in the same way we calculated the effective number of Styles (figure S1).

$$D_T^T = \exp\left(-\sum_{i=1}^{8} \bar{q}_i^T \ln \bar{q}_i^T\right) \tag{4}$$

$$D_T^H = \exp\left(-\sum_{i=1}^{8} \bar{q}_i^H \ln \bar{q}_i^H\right) \tag{5}$$

$$D_T := \frac{D_T^T + D_T^H}{2}, \tag{6}$$

where we define $D_T$ as the overall measure of topic diversity. $D_T$ is shown in the main manuscript (figure 4). The individual H- and T-Topic diversities $D_T^T$ and $D_T^H$ are provided in figure **??**. It is evident that the significant diversity decline in the 1980s is mainly due to a decline in timbral topic diversity, while harmonic diversity shows no sign of sustained decline.

**Disparity.**

In contrast to diversity, disparity corresponds to morphological variety, variety of measurements. Two ecosystems of equal diversity can have different disparity, depending on the extent to which the phenotypes of species differ. A variety of measures, such as average pairwise character dissimilarity and the total variance (sum of univariate variance) [19, 20] have been used to measure disparity. We adopt the square root of total variance, a metric called *total standard deviation* [21, p. 37] as our measure of disparity, i.e. given a set of $N$ observations on $T$ traits as a matrix $X = (x_{n,m})$, we define it as

$$D_Y = \sqrt{\sum_{t=1}^{T} \text{Var}(x_{\cdot, m})}. \tag{7}$$

We apply our disparity measure $D_Y$ to the 14-dimensional matrix of principal components (derived from the topics, as described in Section M.9).

## M.11 Identifying musical revolutions

In order to identify points at which the composition of the charts significantly changes, we employ Foote novelty detection [22], a technique often used in MIR [23]. First we pool the 14-dimensional principal component data (see Section M.9) into quarters by their first entry into the charts (January-March, April-June, July-September, October-December) using the quarterly mean of each principal component. We

then construct a matrix (see figure 5 in main text) of pairwise distances between each quarter. Foote's method consists of convolving such a distance matrix with a so-called checkerboard kernel along the main diagonal of the matrix. Checkerboard kernels represent the stylised case of homogeneity within regions (low values in the upper right and bottom-left quadrants) and dissimilarity between regions (high values in the other two quadrants). In such a situation, i.e. when one homogenous era transitions to another, the convolution results in high values.

A kernel with a half-width of 12 quarters (3 years) compares the 3 years prior to the current quarter to those following the current quarter (figure M5). We follow Foote [22] in using checkerboard kernels with Gaussian taper (standard deviation: 0.4 times the half-width). The kernel matrix entries corresponding to the central, "current" quarter are set to zero.

Many different kernel widths are possible. Figure 5B in the main text shows the novelty score for kernels with half-widths between 4 quarters (1 year) and 50 quarters (12.5 years). We can clearly make out three major 'revolutions' (early 1960s, early 1980s, early 1990s) that result in high novelty scores for a wide range of kernel sizes.

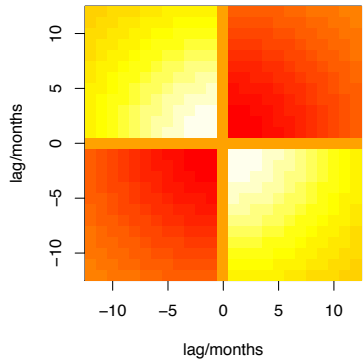

Figure M5: Foote checkerboard kernel for novelty detection.

In order to be able to assess the significance of these regions we compared their novelty scores against novelty values obtained from randomly permuted distance matrices. We first repeated the process 1000 times on distance matrices with randomly permuted quarters. For every kernel size we then calculated the quantiles at confidence levels $\alpha = 0.95$, 0.99 and 0.999. The results are shown as contour lines in figure 5B in the main text.

For further analysis we choose the time scale depicted with a half-width lag of 12 quarters (3 years). This results in three change regions at confidence $p < 0.01$ given in Table ??. The 'revolution' points are the points of maximum Foote novelty within the three regions of significant change, see Table ??. Note that there are no significant changes at small time scales ($< 2$ years). On the other hand, all quarters have significant change at large time scales, i.e. the charts evolve long-term.

## M.12  Identifying Styles that change around each revolution

To identify the styles (clusters) that change around each revolution, we obtained the frequencies of each style for the 24 quarters flanking the peak of a revolution, and estimated the rate of change per annum by a quadratic model. We then used a tag-enrichment analysis to identify those tags associated with each style just around each revolution, see Table S2.

# S  Supplementary Text & Tables

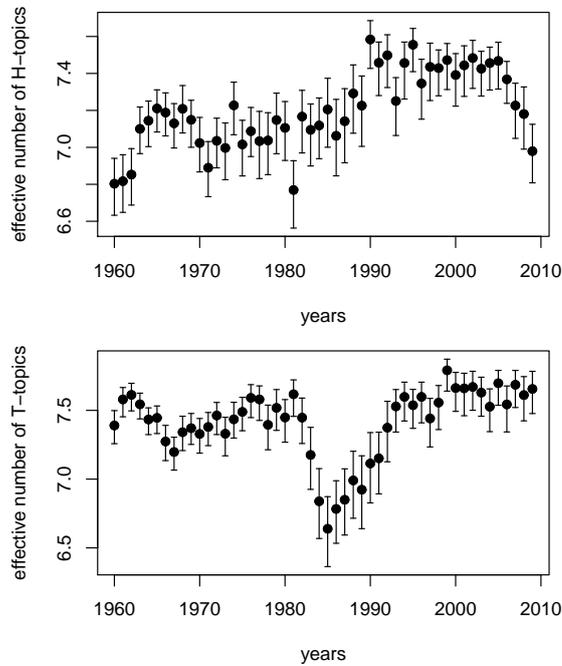

Figure S1: Evolution of Topic diversity. Year-wise topic diversity measures: A  $D_T^T$; B  $D_T^H$.

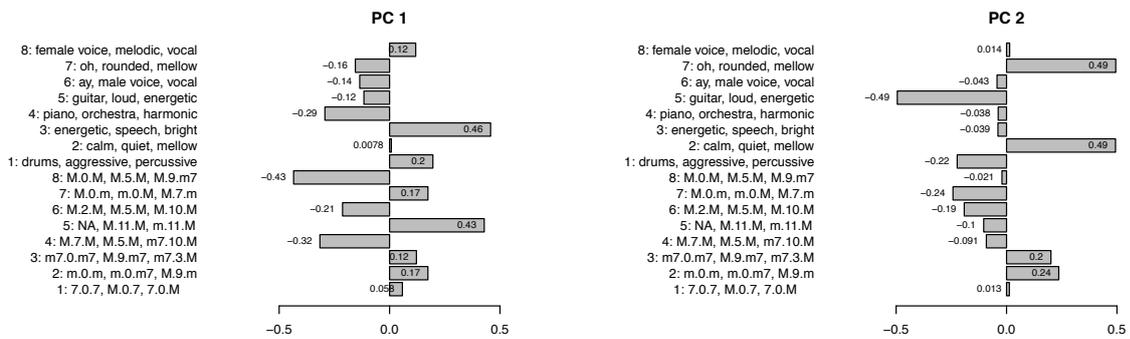

Figure S2: Loadings of the first 2 principal components extracted from the Topic probabilities (see Section M.9).

Table S1: Enrichment analysis: Last.fm user tag over-representation for all Styles over the complete data set (only those with $P < 0.05$)

| Style 1 | Style 2 | Style 3 | Style 4 | Style 5 | Style 6 | Style 7 |
| --- | --- | --- | --- | --- | --- | --- |
| northernsoul | hiphop | easylistening | funk | rock | femalevocal | country |
| soul | rap | country | blues | classicrock | pop | classiccountry |
| hiphop | gangstarap | lovesong | jazz | pop | rnb comb | folk |
| dance | oldschool | piano | soul | newwave | motown | rockabilly |
| rap | dirtysouth | ballad | instrumental | garage | soul | southernrock |
| vocaltrance | dance | classiccountry | rocknroll comb | hardrock | glee | |
| house | dancehall | jazz | northernsoul | garagerock | soundtrack | |
| | westcoast | doowop | rockabilly | british | musical | |
| | party | swing | | | | |
| | comedy | lounge | | | | |
| | reggae | malevocal | | | | |
| | newjackswing | 50s | | | | |
| | classic | singersongwriter | | | | |
| | sexy | romantic | | | | |
| | urban | softrock | | | | |
| | | christmas | | | | |
| | | acoustic | | | | |

| Style 8 | Style 9 | Style 10 | Style 11 | Style 12 | Style 13 |
| --- | --- | --- | --- | --- | --- |
| dance | classicrock | lovesong | funk | soul | rock |
| newwave | country | slowjams | blues | rnb comb | hardrock |
| pop | rock | soul | dance | funk | alternative |
| electronic | singersongwriter | folk | bluesrock | disco | classicrock |
| synthpop | folkrock | rnb comb | newwave | slowjams | alt indi rock comb |
| freestyle | folk | neosoul | electronic | dance | hairmetal |
| rock | pop | femalevocal | synthpop | neosoul | poppunk |
| eurodance | softrock | singersongwriter | hiphop | newjackswing | punkrock |
| newjackswing | | acoustic | hardrock | smooth | punk |
| triphop | | romantic | | oldschoolsoul | poprock |
| funk | | mellow | | | metal |
| | | easylistening | | | powerpop |
| | | jazz | | | emo |
| | | beautiful | | | numetal |
| | | smooth | | | heavymetal |
| | | softrock | | | glamrock |
| | | ballad | | | grunge |
| | | | | | british |

Table S2: Identifying Styles that change around each revolution and the associated over-represented genre tags in the 24 quarters flanking the revolutions.

| revo-lution | style cluster | estim. (linear) | p | estim. (quad.) | p | tags |
|---|---|---|---|---|---|---|
| 1964 | 1 | 0.044 | 0.019 | 0.050 | 0.009 | northernsoul, motown, soul, easylistening, jazz |
|  | 2 | -0.014 | 0.291 | 0.008 | 0.528 | comedy, funny, jazz, easylistening |
|  | 3 | -0.177 | 0.000 | 0.014 | 0.631 | easylistening, jazz, swing, lounge, doowop |
|  | 4 | -0.060 | 0.118 | -0.009 | 0.806 | jazz, northernsoul, soul, instrumental, blues |
|  | 5 | 0.060 | 0.014 | -0.041 | 0.081 | garagerock, garage, british, psychedelic |
|  | 6 | -0.171 | 0.000 | -0.045 | 0.282 | femalevocal, motown, northernsoul, soul, doowop |
|  | 7 | -0.018 | 0.495 | -0.003 | 0.908 | british, folk, surf, malevocal, rocknroll comb |
|  | 8 | 0.062 | 0.000 | 0.010 | 0.441 | garagerock, instrumental, northernsoul, surf, soul |
|  | 9 | 0.083 | 0.014 | 0.006 | 0.861 | rocknroll comb, northernsoul, motown, soul, garage |
|  | 10 | -0.020 | 0.349 | 0.034 | 0.109 | folk, easylistening, jazz, swing |
|  | 11 | 0.037 | 0.131 | -0.022 | 0.349 | garagerock, blues, soul, northernsoul, instrumental |
|  | 12 | 0.043 | 0.001 | 0.000 | 0.996 | northernsoul, soul, motown |
|  | 13 | 0.121 | 0.000 | -0.005 | 0.821 | psychedelic, garagerock, psychedelicrock, motown, british |
| 1983 | 1 | 0.031 | 0.350 | -0.050 | 0.136 | newwave, synthpop, disco |
|  | 2 | 0.024 | 0.189 | 0.010 | 0.569 | oldschool, funk, comedy |
|  | 3 | -0.112 | 0.000 | 0.021 | 0.426 | lovesong, softrock, easylistening, romantic |
|  | 4 | -0.022 | 0.448 | -0.005 | 0.846 | newwave, disco, progressiverock |
|  | 5 | 0.020 | 0.614 | -0.019 | 0.617 | newwave, rock, classicrock, progressiverock, pop, synthpop |
|  | 6 | -0.013 | 0.598 | -0.005 | 0.828 | femalevocal, disco, pop, reggae, musical, soundtrack |
|  | 7 | -0.098 | 0.003 | 0.015 | 0.598 | newwave, classiccountry, softrock, newromantic, rock, southernrock |
|  | 8 | 0.173 | 0.000 | -0.029 | 0.190 | newwave, pop, rock, disco, synthpop |
|  | 9 | -0.065 | 0.057 | 0.065 | 0.055 | classicrock, rock, softrock, progressiverock, newwave, pop |
|  | 10 | -0.047 | 0.110 | -0.004 | 0.882 | lovesong, softrock |
|  | 11 | 0.048 | 0.089 | -0.058 | 0.043 | newwave, synthpop, rock, classicrock, hardrock |
|  | 12 | -0.095 | 0.001 | 0.011 | 0.664 | funk, disco, soul, smoothjazz, dance |
|  | 13 | 0.150 | 0.000 | 0.034 | 0.236 | rock, classicrock, hardrock, newwave, progressiverock, southernrock, powerpop, heavymetal, hairmetal |
| 1991 | 1 | 0.034 | 0.161 | -0.018 | 0.457 | house, freestyle, synthpop, newwave, electronic, dance |
|  | 2 | 0.325 | 0.000 | 0.043 | 0.187 | hiphop, rap, oldschool, newjackswing, gangstarap, eurodance, westcoast, dance |
|  | 3 | 0.056 | 0.011 | 0.011 | 0.596 | ballad, lovesong |
|  | 4 | 0.005 | 0.822 | -0.034 | 0.153 | dance, hairmetal, hardrock, metal, newjackswing |
|  | 5 | -0.085 | 0.003 | 0.003 | 0.901 | rock, hardrock, hairmetal, pop, freestyle |
|  | 6 | 0.000 | 0.999 | 0.015 | 0.344 | femalevocal, pop, slowjams, dance, rnb comb, ballad |
|  | 7 | 0.023 | 0.307 | 0.007 | 0.742 | rock, hardrock, hairmetal |
|  | 8 | -0.023 | 0.584 | -0.023 | 0.576 | dance, newjackswing, freestyle, pop, electronic, synthpop, newwave, australian |
|  | 9 | -0.015 | 0.690 | 0.046 | 0.224 | rock, pop, softrock, ballad, hardrock |
|  | 10 | 0.042 | 0.235 | 0.008 | 0.813 | slowjams, lovesong, rnb comb |
|  | 11 | -0.065 | 0.016 | 0.030 | 0.248 | newwave, dance, synthpop, hairmetal, rock, freestyle |
|  | 12 | -0.035 | 0.447 | -0.017 | 0.712 | newjackswing, rnb comb, dance, slowjams, house, pop, soul |
|  | 13 | -0.207 | 0.000 | -0.112 | 0.023 | hardrock, hairmetal, rock, classicrock, metal, heavymetal, thrashmetal, madchester |

13



\end{document}